\documentclass[conference]{IEEEtran}
\IEEEoverridecommandlockouts
\usepackage{cite}
\usepackage{amsmath,amssymb,amsfonts}
\usepackage{algorithmic}
\usepackage{graphicx}
\usepackage{textcomp}
\usepackage{xcolor}
\def\BibTeX{{\rm B\kern-.05em{\sc i\kern-.025em b}\kern-.08em
    T\kern-.1667em\lower.7ex\hbox{E}\kern-.125emX}}

\usepackage{url}
\usepackage{soul} 

\newcommand\nvidia{NVIDIA}
\newcommand\netherlands{\textit{F3 Netherlands}}
\newcommand\parihaka{\textit{Parihaka}}
\newcommand\kahu{\textit{Kahu}}
\newcommand\kmeans{\textit{k}-means}

\begin{document}

\title{Accelerating Multi-attribute Unsupervised Seismic Facies Analysis With RAPIDS}

\renewcommand{\thefootnote}{\fnsymbol{footnote}} 

\author{
\IEEEauthorblockN{Ot\'{a}vio O. Napoli}
\IEEEauthorblockA{\textit{Institute of Computing} \\
\textit{UNICAMP}\\
Campinas, Brazil \\
onapoli@lmcad.ic.unicamp.br}
\and
\IEEEauthorblockN{Vanderson Martins do Rosario}
\IEEEauthorblockA{\textit{Institute of Computing} \\
\textit{UNICAMP}\\
Campinas, Brazil \\
vrosario@lmcad.ic.unicamp.br }
\and
\IEEEauthorblockN{Jo\~{a}o Paulo Navarro}
\IEEEauthorblockA{\textit{NVIDIA} \\
\textit{NVIDIA}\\
São Paulo, Brazil \\
jpnavarro@nvidia.com}
\and
\IEEEauthorblockN{Pedro M\'{a}rio Cruz e Silva}
\IEEEauthorblockA{\textit{NVIDIA} \\
\textit{NVIDIA}\\
Rio de Janeiro, Brazil \\
pcruzesilva@nvidia.com}
\and
\IEEEauthorblockN{Edson Borin}
\IEEEauthorblockA{\textit{Institute of Computing} \\
\textit{UNICAMP}\\
Campinas, Brazil \\
borin@unicamp.br}
}


\maketitle

\begin{abstract}
Classification of seismic facies is done by clustering seismic data samples based on their attributes. Year after year, 3D datasets used by exploration geophysics increase in size, complexity, and number of attributes, requiring a continuous rise in the classification performance. In this work, we explore the use of Graphics Processing Units (GPUs) to perform the classification of seismic surveys using the well-established Machine Learning (ML) method \kmeans. We show that the high-performance distributed implementation of the \kmeans\ algorithm available at the RAPIDS library can be used to classify facies in large seismic datasets much faster than a classical parallel CPU implementation (up to 258-fold faster in \nvidia\ V100 GPUs), especially for large seismic blocks.  We tested the algorithm with different real seismic volumes, including \netherlands, \parihaka, and \kahu\ (from 12GB to 66GB). 
\end{abstract}
\vspace{-0.2cm}

\section{Introduction}

Lithofacies classification is a crucial step in seismic interpretation. The accurate classification of samples in terms of similarities leads to a better understanding of the areas of interest. Subsurface events contain valuable spatio-temporal information that has scientific and commercial importance, making accurate and fast interpretations a competitive advantage in exploration geophysics. Formerly in seismic analysis, this process consisted of assigning lithofacies manually by human interpreters, following the amplitude responses. 
This labor-intensive task is being consistently improved by the use of automatic and semi-automatic interpretation tools. 
However, with the increasing quality of acquisition sensors, the size of 3D seismic surveys is facing significant improvement in terms of definition, which ultimately leads to longer processing times. 
Furthermore, the interpretation quality may expressively be improved by adding new perspectives to the original amplitude volume, using, for example, derived attributes such as phase, frequency, and envelope. 
However, in the context of multi-attribute analysis, the computational complexity and requirements proportionally increase.

In the modern era of computing, massively parallel architectures are central in high-performance computing. 
The use of Graphics Processing Units (GPUs) at scalable infra-structures is enabling the processing of these increasing datasets. 
GPUs are well-known in the geophysics domain and are commonly used to solve inversion and migration problems such as reverse time migration~\cite{gpu_rtm}, full-waveform inversion~\cite{fwi}, Kirchhoff migration~\cite{Kirchhoff}, and least-squares migration~\cite{lqm}, for instance.
We also find in literature efforts on porting attributes computations to GPUs as the work proposed by \cite{curvature}. In their paper, an interactive-time curvature estimate was achieved maximizing the memory access pattern and loading data to GPU shared memory using a circular buffer.

Data-driven seismic attributes are well explored by the geophysics community. 
Unsupervised learning approaches may be used to categorize waveforms, or volume samples, in classes (clusters). 
The intuition behind is to use established algorithms such as Principal Component Analysis (PCA)~\cite{pca}, \kmeans~\cite{Lloyd_1982} or Self-organizing Maps (SOM)~\cite{kohonen1982self} to automatically find the data distribution and properly classify the samples. 
Thus, similar feature vectors tend to be at the same cluster, indicating they have similar expressions~\cite{cho_mar}. 
For each volume of interest, we first train the algorithm and then perform predictions to classify the samples. 
The volume size has a direct impact on the attribute training time, meaning that very large datasets may become unfeasible to compute if we do not apply scalable parallelization strategies. 
For this reason, we present in this work a study exploring the large-scale implementation of \kmeans\ clustering available at the open GPU data science library named RAPIDS\footnote{\url{https://rapids.ai/}}. 
This package provides efficient ML algorithms running on multiple NVIDIA GPUs. 
With RAPIDS, we are able to train \kmeans\ over 3D seismic volumes up to 258x faster than conventional CPU versions.

The contributions of this work are summarized as follows:

\begin{itemize}
    \item We tested and compared two open-source implementations of multi-attribute unsupervised seismic facies classification using \kmeans. 
    One with a DASK-based \cite{dask} CPU implementation and other with a RAPIDS GPU-based implementation.  
    
    \item We tested and analyzed the efficiency of the CPU \kmeans\ implementation approach available in DASK-ML\footnote{https://ml.dask.org/} in a 40-threads computer node, comparing it with the RAPIDS implementation for 2, 4, 8 and 16 \nvidia\ V100 GPUs. 
    The tests were performed using three seismic datasets (\netherlands, \parihaka, and \kahu)\footnote{\url{https://wiki.seg.org/wiki/Open_data}} and we were able to show a speedup of 258x with RAPIDS over the CPU baseline approach.
\end{itemize}

The remaining of the text is organized as follows. 
We first describe the seismic facies classification problem, the algorithms that can be used, the pros and cons of \kmeans\, and some of its implementations in the \textit{Seismic Facies Classification} Section. 
Then, in \textit{ML Workflow} Section, we detail how we use the RAPIDS \kmeans\ implementation to perform seismic data classification. 
The \textit{Experimental Setup} Section describes the computational and data resources that we used to perform our experiments. 
The \textit{Experimental Results} Section presents and discusses the results we obtain followed by conclusions.
\vspace{-0.2cm}
\section{Seismic Facies Classification}
\label{sec:seismicclassification}

Seismic facies classification is the problem of assigning specific classes to samples of seismic volumes based on their attributes. This classification allows the visualization of different geological settings, demanding complex analysis of enormous amounts of data. 
To handle the challenge of interpreting increasingly larger datasets, the use of ML has become an important tool. Unsupervised ML techniques are now widely used for this propose aiming to find natural clusters among different attributes that better highlight seismic patterns such as variances in amplitude and steeply dipping, low amplitude dipping areas, continuous dipping reflectors, among others. The work proposed by T. Zhao et.al \cite{zhao2015comparison} compares a set of unsupervised learning techniques used to classify seismic facies including Self-Organizing Maps (SOM), \kmeans\, and Neural Networks (NN). 

The use of unsupervised learning yielded good results in several seismic interpretation tasks. In general, these techniques benefit from fast training and prediction with lower computational resources requirements than complex learning algorithms, such as Deep Learning (DL). However, the data size still growing either by new acquisition methods, new formats, and storage technologies or by new attributes that become relevant for analysis. This increases the computing needs, even for low complexity ML methods. TerraNubis\footnote{\url{https://terranubis.com/}}, for instance, shows seismic datasets at a 200GB scale, which may be unfeasible to process, extract features, and to train with conventional tools (e.g. scikit-learn~\cite{sklearn_api}).

\textbf{SOM}. Also named Kohonen maps and developed by T. Kohonen \cite{kohonen1982self} is an NN used to produce a low-dimensional, discretized representation of the input space. 
SOM applies competitive learning instead of using error-correction learning approaches (e.g. backpropagation), preserving the properties of the input space. This last property is very useful in the visualization task, especially because similar clusters end adjacent to each other in the latent space. 
SOM is widely used in seismic facies classification works~\cite{roy2010automatic,zhao2015comparison,song2017multi,zhao2018seismic}.

\textbf{\kmeans}. Is an iterative clustering algorithm that minimizes the sum of distances from all points in a dataset to its clusters' centroids. 
That is, \kmeans\ tries to minimize the following equation: 
\vspace{-0.1cm}
\begin{equation}
E = \sum_{i=1}^{k}\sum_{j=1}^{n_{i}} d^{2}(x_{ij},m_{i}),
\label{eqn:kmeans}
\end{equation}
where $E$ represents the sum of the euclidian distance from all samples in the dataset to the centroids, $x_{ij}$ is the $j$th sample in the $i$th cluster, $m_{i}$ is the center or mean of the $i$th cluster, $n_{i}$ is the number of samples in the cluster, $k$ is the number of clusters and $d$ is the euclidian distance which is defined by:

\begin{equation}
d^{T}(x_{ij}, m{i}) = (x_{ij} - m_{i})*(x_{ij} - m_{i})^{T}.
\label{eqn:euclidean}
\end{equation}

For each iteration, the centroids are moved towards the data center of mass and the algorithm stops when reaching a given number of iterations or distances sum ($E$) reaches a defined convergence tolerance.
In the end, the result is a set of clusters defined by the closest centroid that partitioning the data as well as possible.

In contrast to SOM, the number of clusters in \kmeans\ is a hyparameter that needs to be set before the training phase. Moreover, \kmeans\ clustering has no structure (and con\-se\-quent\-ly no relationship in the cluster numbering) usually generating a not ideal seismic facies visualization. 
In practice, \kmeans\ is normally applied to estimate the correct number of clusters in the data while SOM is used to generate the clusters for final visualization. 
However, despite its limitations, \kmeans\ is very robust and widely available in several ML libraries and much faster than SOM when handling large amounts of data.
Thus, with the growth of seismic datasets resolution, dimensionality reduction techniques and faster clustering algorithms are central in modern interpretation platforms.

We explore and present how \kmeans\ can be used to quickly obtain classifications on seismic volumes in \nvidia\ GPUs using the RAPIDS implementation with single and multiple devices. 
\kmeans\ shows not only to be fast in single GPUs, but also scalable running in multi-GPU servers.   

\subsection{\kmeans\ Implementations}

Sabeti and Javaherian~\cite{sabeti2009seismic} use \kmeans\ on multi-attribute 3D seismic data, showing that the classifier is useful to extract information about underground beddings and lateral changes in layers.
Some works have presented parallel and efficient implementations of \kmeans\ for CPUs and accelerators, even for geophysics purposes. 
For instance, \cite{zhao2015comparison} uses a \kmeans\ version implemented with MPI to accelerate the clustering with multiple-CPU nodes. \cite{huang2017scalable} improves this approach by presenting their design of a scalable seismic analytics platform built upon Apache Spark and its managing features to process and visualize seismic data using GPU-accelerated nodes. 
\cite{huang2017scalable} employ their framework to identify geologic faults, but using deep learning.

In 2018, an open-source consortium named RAPIDS was announced. RAPIDS is a suite of data processing and ML libraries that enables efficient data-science back-ended by GPU acceleration. 
It has a great advantage of providing a user-friendly abstraction to well-known data-science libraries such as Numpy~\cite{numpy} and Pandas~\cite{pandas}, as well as exposing high-level GPU parallelism and high-bandwidth memory speed trough CUDA~\cite{cuda}.
With RAPIDS, we employed DASK to allow distributed processing with multiple GPUs.
As far as we know, this is the first work to explore the performance and scalability of RAPIDS applied to seismic facies classification.

To evaluate the \kmeans\ performance, we used an ML workflow similar to~\cite{zhao2015comparison,zhao2018seismic}. The workflow is detailed in the next section.
\vspace{-0.2cm}
\section{ML Workflow}
\label{sec:mlworkflow}

\begin{figure}[!htb]
    \centering
    \includegraphics[scale=0.82]{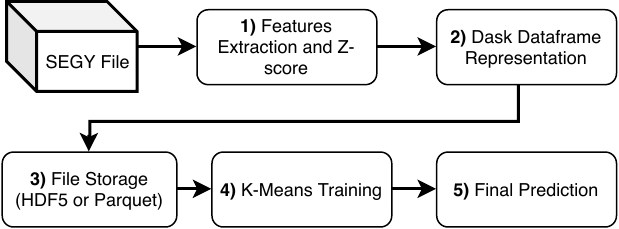}
    \caption{ML workflow}
    \label{fig:seismic-workflow}
\end{figure}

In Figure~\ref{fig:seismic-workflow} we depict our workflow. All steps run using DASK~\cite{dask}, an open-source and flexible library for Python distributed computing.
DASK allows easy creations and extensions of computing clusters.

Given a SEGY cube as input, we perform the feature extraction (1) using \texttt{d2geo}, an open-source Python library for computing seismic attributes built on top of DASK parallelization suite~\cite{d2geo}.
Many attributes are well suited for facies classification since most of them highlight continuities, faults, impedance contrasts, among other important characteristics. 
Wrona \textit{et al.}~\cite{wrona2018seismic} lists several attributes that allow accurate predictions of facies, mostly based on non-linear transforms, such as the Hilbert Transformation. 
We used the following set of attributes as features in our experiments: \textit{Amplitude, Cosine Instantaneous Phase, Dominant Frequency, Envelope, Instantaneous Bandwidth, Instantaneous Frequency, Instantaneous Phase, Reflection Intensity, and Second Derivative}.  
After feature extraction, values are arranged in a DASK dataframe (2), i.e., a 2D matrix where each line corresponds to a specific point of the seismic cube and each column is the attribute value for that point. 
For classification purposes, we consider each line of the dataframe as a sample with their respective feature vector (nine column). Afterward, we normalized the dataframe column values using the \textit{z}-score technique.

We convert and store (3) the dataframe in a Parquet file~\cite{apache-parquet}, an open-source Apache-designed format that stores nested data structures in a flat columnar way.
Parquet is designed to handle large amounts of data and to support parallel access, which is desired when dealing with distributed computing. 

In the fourth step (classification) we use for comparison two \kmeans\ implementations: one provided by DASK-ML library, which allows parallel training in multiple CPUs, and other available by the RAPIDS cuMLlibrary~\cite{rapids-cuml}, which also uses DASK to support multi-GPUs training. 
Both offer similar interfaces to the well-known Python library for ML: scikit-learn~\cite{scikit-learn}. Both implement the \kmeans\ algorithm with average time complexity of $O(k n t)$, where $k$ is the number of clusters, $n$ is the number of samples and $t$ is the number of iterations. 

During the classification step, the input data (Parquet format) is not loaded into the memory beforehand. Instead, it is loaded on-demand, allowing the training process to overlap data ingestion with other operations. This process is orchestrated by DASK and is very useful when dealing with distributed memory. 
The training results generate the classification (5), which is also used for visualization - one integer value per class.

We used the proposed workflow in our experiments to test both \kmeans\ implementations. The experiments' scope, infrastructure, and datasets are described in the next section. 

\vspace{-0.2cm}
\section{Experimental Setup}
\label{sec:setup}
\vspace{-0.1cm}
\subsection{3D Seismic Datasets}

For tests, we use three seismic post-stacked open datasets, being them: (a) \netherlands\ seismic survey, which is a small 3D marine data from offshore of Netherlands; (b) \parihaka, a marine data from New Zealand and; (c) \kahu, also a marine data from New Zealand. 
The original dataset sizes before and after feature extraction and conversion are shown in Table~\ref{tbl:data-sizes}.

\begin{table}[!htb]
    \scriptsize
    \caption{Seismic surveys and dataframe sizes.}
    \begin{center}
        \begin{tabular}{lcc}
        \hline
        \textbf{Data} & \textbf{Original Data Size} & \textbf{Feature Extracted Dataframe Size} \\ \hline
        \netherlands\  & 1.3GB                & 12.5GB                  
        \\ 
        \parihaka\        & 3.9GB                & 30.2GB                  \\
        \kahu\            & 6.1GB                & 66.6GB                  \\ \hline
        \end{tabular}
    \end{center}
    \label{tbl:data-sizes}
\end{table}

\subsection{Hardware and Software Infrastructure}


Our experiments were performed on the hardware specified in Table \ref{tbl:infrastructure}. Experiments with 2 and 4 GPUs were executed on a DGX Station, 8 GPUs on DGX-1, and 16 GPUs on DGX-2. We used the RAPIDS 0.13 docker container image, which includes all \nvidia\ libraries.

\begin{table}[!htb]
    \scriptsize
    \caption{Infrastructure used in the experiments.}
    \begin{center}
        \begin{tabular}{p{1cm}lcll}
        \hline
        \textbf{Node}                                              & \textbf{Processor}                                                       & \textbf{Cores} & \textbf{RAM} & \textbf{GPUS}                                                    \\ \hline
        \begin{tabular}[c]{@{}l@{}}DGX Station\end{tabular} & \begin{tabular}[c]{@{}l@{}}Xeon E5-2698 v4\end{tabular}          & 20             & 256GB        & \begin{tabular}[c]{@{}l@{}}4x Tesla V100 - 32GB\end{tabular}  \\ 
        DGX-1                                                  & \begin{tabular}[c]{@{}l@{}}Xeon E5-2698 v4\end{tabular}          & 20             & 512GB        & \begin{tabular}[c]{@{}l@{}}8x Tesla V100 - 32GB\end{tabular}  \\ 
        DGX-2                                                  & \begin{tabular}[c]{@{}l@{}}Dual Xeon  Platinum 8168\end{tabular} & 24             & 1.5TB        & \begin{tabular}[c]{@{}l@{}}16x Tesla V100 - 32GB\end{tabular} \\ \hline
        \end{tabular}
    \end{center}
    \label{tbl:infrastructure}
\end{table}

\section{Experimental Results}
\label{sec:results}

\subsection{\netherlands\ Seismic Data}

\begin{figure}[!htb]
    \centering
    \includegraphics[scale=0.27]{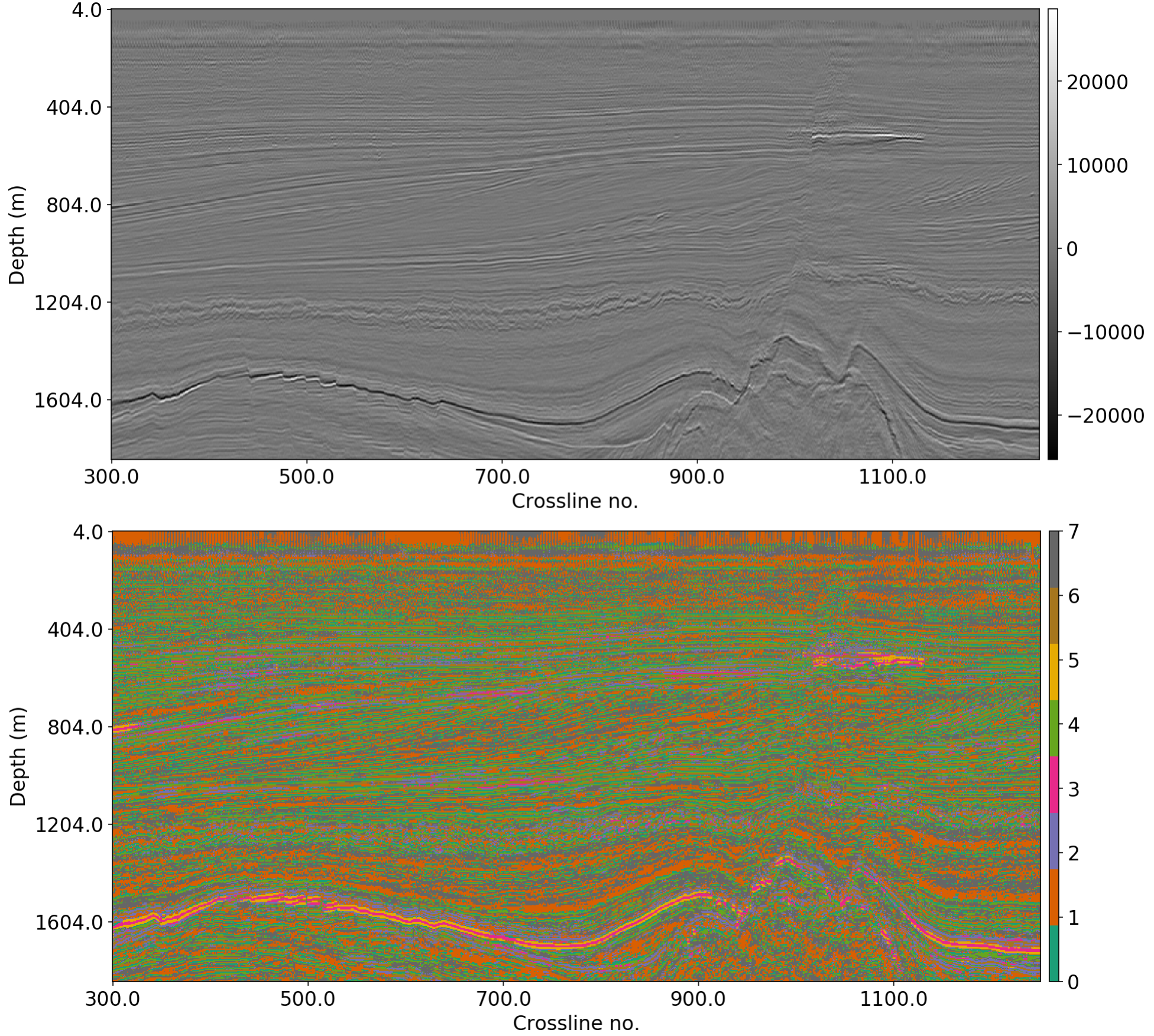}
    \caption{\netherlands\ inline 100 (top). Results for \kmeans\ prediction, using 8 clusters (bottom).}
    \label{fig:f3-inline-100-classify-1}
\end{figure}

Figure~\ref{fig:f3-inline-100-classify-1} (top) shows a section of the \netherlands\ (inline 100). We executed \kmeans\ varying the number of clusters from 5 to 12. 
The result of the facies classification using 8 clusters can be seen on bottom of Figure~\ref{fig:f3-inline-100-classify-1}, where the colors indicate the associated facies.
Based on the features used, it is possible to note that the classification highlights continuous, horizontal, and low amplitude reflectors with the green color (face 4).
Very high amplitude reflectors are highlighted with strong purple and yellow colors (facies 2 and 3) while the orange and gray ones (facies 1 and 7) denote continuous oblique areas.

\begin{figure}[!htb]
    \centering
    \includegraphics[scale=1.5]{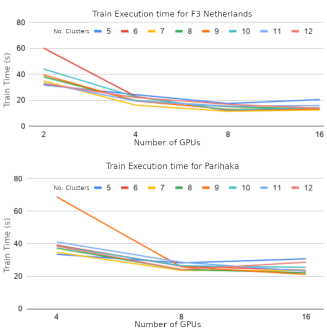}
    \caption{Train execution time of \netherlands\ (top) and \parihaka\ (bottom) datasets using different number of GPUs. Colors represent different number of clusters.}
    \label{fig:gpu-train}
\end{figure}

Figure~\ref{fig:gpu-train} shows training performance using multiple GPUs with RAPIDS. 
It is worth notice that performance improves by 1.88x when we increase the number of GPUs from 2 to 4, but there is a marginal (~3\%) performance reduction when increasing the number of GPUs from 8 to 16. 
This occurs due to low GPU occupancy (memory and CUDA cores utilization) when processing small datasets. We also notice that the performance does not significantly change when using a different number of clusters, as the \kmeans\ time complexity grows linearly with the number of clusters. In fact, the time complexity of the implementations is $O(nkt)$, where $n$ is the number of samples, $k$ is the number of clusters and $t$ is the number of iterations.

A speedup comparison with CPU DASK implementation is shown in Figure \ref{fig:cpu-speedup} (blue). The speedup is the geometric mean of all speedups from GPU over CPU using 5 to 12 \kmeans\ clusters. The highest speedup was about 186-fold when using 8 GPUs compared to DASK \kmeans\ CPU implementation with 40 threads. The overall speedup does not increase when using 16 GPUs.

\subsection{\parihaka\ Seismic Data}

\begin{figure}[!htb]
    \centering
    \includegraphics[scale=1.5]{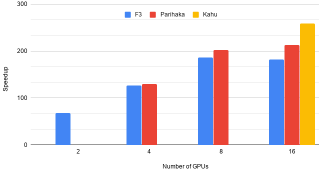}
    \caption{RAPIDS speedup compared to CPU DASK implementation using 40 threads.}
    \label{fig:cpu-speedup}
\end{figure}

For \parihaka, we also performed training procedures ranging from 5 to 12 clusters. 
At the time of our experimentation, the default RAPIDS \kmeans\ memory footprint was up to 4 times the size of the input data. Thus, as shown in Table~\ref{tbl:data-sizes} (\parihaka\ dataframe), we cropped the dataset from inline 0 to 700 summing a total of 30.2GB, so it could fit the available GPU memory. 
The execution times for training is shown in Figure~\ref{fig:gpu-train} (bottom). Similarly to \netherlands, the execution time decreases as the number of GPUs increases from 4 to 8 (about 55\%). 
However, the processing time is almost the same when increasing the number of GPUs from 8 to 16, for the same reason as the previous dataset. 
The speedup achieved was about 217-fold, as depicted in Figure~\ref{fig:cpu-speedup} (red bars).


\subsection{\kahu\ Seismic Data}

For \kahu\ seismic data we performed training using 16 GPUs, in order to fit the training data to the available GPU memory.
The speedup over the CPU implementation is presented in Figure~\ref{fig:cpu-speedup} (yellow). With a larger dataset, we were able to saturate the GPU resources, achieving a training speedup of 258-fold. Comparing to the speedup of \textit{Parihaka}, which is nearly half of the size of \kahu, we had a speedup of 18\%.

\vspace{-0.2cm}
\section{Conclusions}
\label{sec:conclusion}

Seismic facies classification is an important step in seismic interpretation, allowing the visualization of different geological settings. We presented a scalable ML workflow to support seismic facies classification over large 3D seismic volumes. We use the RAPIDS library and compared their \kmeans\ implementation with a classical multi-core CPU implementation. The results show a speedup up to 258-fold in a DGX-2 server, with 16x GPUs. RAPIDS \kmeans\ offer all necessary tools to easily scale training when GPU resources are available. However, we verified that training in small datasets regimes have a gain upper limit, meaning that adding more GPU devices may not proportionally reduce training time. This pipeline may be adapted to several seismic processing and interpretation tasks, aiming to provide efficient and scalable solutions to different applications.

\section{ACKNOWLEDGMENTS}
We would like to thank \nvidia\ for the computational resources used in our experiments and Petrobras for research support.

\bibliographystyle{IEEEtran}
\bibliography{bibliography}

\begin{thebibliography}{10}
\providecommand{\url}[1]{#1}
\csname url@samestyle\endcsname
\providecommand{\newblock}{\relax}
\providecommand{\bibinfo}[2]{#2}
\providecommand{\BIBentrySTDinterwordspacing}{\spaceskip=0pt\relax}
\providecommand{\BIBentryALTinterwordstretchfactor}{4}
\providecommand{\BIBentryALTinterwordspacing}{\spaceskip=\fontdimen2\font plus
\BIBentryALTinterwordstretchfactor\fontdimen3\font minus
  \fontdimen4\font\relax}
\providecommand{\BIBforeignlanguage}[2]{{%
\expandafter\ifx\csname l@#1\endcsname\relax
\typeout{** WARNING: IEEEtran.bst: No hyphenation pattern has been}%
\typeout{** loaded for the language `#1'. Using the pattern for}%
\typeout{** the default language instead.}%
\else
\language=\csname l@#1\endcsname
\fi
#2}}
\providecommand{\BIBdecl}{\relax}
\BIBdecl

\bibitem{gpu_rtm}
J.~{Cabezas}, M.~{Araya-Polo}, I.~{Gelado}, N.~{Navarro}, E.~{Morancho}, and
  J.~M. {Cela}, ``High-performance reverse time migration on gpu,'' in
  \emph{2009 International Conference of the Chilean Computer Science Society},
  2009, pp. 77--86.

\bibitem{fwi}
\BIBentryALTinterwordspacing
Z.~Wang, J.~Huang, P.~Yong, and Z.~Yang, ``3d variable-grid full waveform
  inversion accelerated on gpu,'' vol. 2019, no.~1, pp. 1--5, 2019. [Online].
  Available:
  \url{https://www.earthdoc.org/content/papers/10.3997/2214-4609.201901001}
\BIBentrySTDinterwordspacing

\bibitem{Kirchhoff}
\BIBentryALTinterwordspacing
G.~Liu, Z.~Yu, J.~Wang, and B.~Li, ``Accelerating kirchhoff pre-stack depth
  migration on a gpu by overlapping ray tracing and imaging,'' \emph{Computers
  \& Geosciences}, vol. 128, pp. 79 -- 86, 2019. [Online]. Available:
  \url{http://www.sciencedirect.com/science/article/pii/S0098300418307519}
\BIBentrySTDinterwordspacing

\bibitem{lqm}
H.~Knibbe, C.~Vuik, and K.~Oosterlee, ``Reduction of computing time for
  least-squares migration based on the helmholtz equation by graphics
  processing units,'' \emph{Computational Geosciences}, vol.~20, no.~2, pp.
  297--315, Jan. 2016.

\bibitem{curvature}
L.~{Martins}, M.~A. {Goncalves da Silva}, M.~{Arruda}, J.~{Duarte}, P.~M.
  {Silva}, R.~{Beauclair Seixas}, M.~{Gattass}, P.~{Souza}, and J.~{Panetta},
  ``Accelerating curvature estimate in 3d seismic data using gpgpu,'' in
  \emph{2014 IEEE 26th International Symposium on Computer Architecture and
  High Performance Computing}, 2014, pp. 105--111.

\bibitem{pca}
\BIBentryALTinterwordspacing
I.~Jolliffe, \emph{Principal Component Analysis}.\hskip 1em plus 0.5em minus
  0.4em\relax Berlin, Heidelberg: Springer Berlin Heidelberg, 2011, pp.
  1094--1096. [Online]. Available:
  \url{https://doi.org/10.1007/978-3-642-04898-2_455}
\BIBentrySTDinterwordspacing

\bibitem{Lloyd_1982}
\BIBentryALTinterwordspacing
S.~Lloyd, ``Least squares quantization in {PCM},'' \emph{{IEEE} Transactions on
  Information Theory}, vol.~28, no.~2, pp. 129--137, mar 1982. [Online].
  Available: \url{https://doi.org/10.1109%2Ftit.1982.1056489}
\BIBentrySTDinterwordspacing

\bibitem{kohonen1982self}
T.~Kohonen, ``Self-organized formation of topologically correct feature maps,''
  \emph{Biological cybernetics}, vol.~43, no.~1, pp. 59--69, 1982.

\bibitem{cho_mar}
S.~Chopra, K.~Marfurt, and R.~Sharma, ``Unsupervised machine learning facies
  classification in the delaware basin and its comparison with supervised
  bayesian facies classification,'' \emph{SEG Technical Program Expanded
  Abstracts}, pp. 2619--2623, 2019.

\bibitem{dask}
\BIBentryALTinterwordspacing
{Dask Development Team}, \emph{Dask: Library for dynamic task scheduling},
  2016. [Online]. Available: \url{https://dask.org}
\BIBentrySTDinterwordspacing

\bibitem{zhao2015comparison}
T.~Zhao, V.~Jayaram, A.~Roy, and K.~J. Marfurt, ``A comparison of
  classification techniques for seismic facies recognition,''
  \emph{Interpretation}, vol.~3, no.~4, pp. SAE29--SAE58, 2015.

\bibitem{sklearn_api}
L.~Buitinck, G.~Louppe, M.~Blondel, F.~Pedregosa, A.~Mueller, O.~Grisel,
  V.~Niculae, P.~Prettenhofer, A.~Gramfort, J.~Grobler, R.~Layton,
  J.~VanderPlas, A.~Joly, B.~Holt, and G.~Varoquaux, ``{API} design for machine
  learning software: experiences from the scikit-learn project,'' in \emph{ECML
  PKDD Workshop: Languages for Data Mining and Machine Learning}, 2013, pp.
  108--122.

\bibitem{roy2010automatic}
A.~Roy, M.~Matos, and K.~J. Marfurt, ``Automatic seismic facies classification
  with kohonen self organizing maps—a tutorial,'' \emph{Geohorizons Journal
  of Society of Petroleum Geophysicists}, vol.~15, pp. 6--14, 2010.

\bibitem{song2017multi}
C.~Song, Z.~Liu, Y.~Wang, X.~Li, and G.~Hu, ``Multi-waveform classification for
  seismic facies analysis,'' \emph{Computers \& Geosciences}, vol. 101, pp.
  1--9, 2017.

\bibitem{zhao2018seismic}
T.~Zhao, F.~Li, and K.~J. Marfurt, ``Seismic attribute selection for
  unsupervised seismic facies analysis using user-guided data-adaptive
  weights,'' \emph{Geophysics}, vol.~83, no.~2, pp. O31--O44, 2018.

\bibitem{sabeti2009seismic}
H.~Sabeti and A.~Javaherian, ``Seismic facies analysis based on k-means
  clustering algorithm using 3d seismic attributes,'' in \emph{Shiraz 2009-1st
  EAGE International Petroleum Conference and Exhibition}.\hskip 1em plus 0.5em
  minus 0.4em\relax European Association of Geoscientists \& Engineers, 2009,
  pp. cp--125.

\bibitem{huang2017scalable}
L.~Huang, X.~Dong, and T.~E. Clee, ``A scalable deep learning platform for
  identifying geologic features from seismic attributes,'' \emph{The Leading
  Edge}, vol.~36, no.~3, pp. 249--256, 2017.

\bibitem{numpy}
T.~E. Oliphant, \emph{A guide to NumPy}.\hskip 1em plus 0.5em minus 0.4em\relax
  Trelgol Publishing USA, 2006, vol.~1.

\bibitem{pandas}
W.~McKinney \emph{et~al.}, ``Data structures for statistical computing in
  python,'' in \emph{Proceedings of the 9th Python in Science Conference}, vol.
  445.\hskip 1em plus 0.5em minus 0.4em\relax Austin, TX, 2010, pp. 51--56.

\bibitem{cuda}
S.~Cook, \emph{CUDA Programming: A Developer’s Guide to Parallel Computing
  with GPUs}, 1st~ed.\hskip 1em plus 0.5em minus 0.4em\relax San Francisco, CA,
  USA: Morgan Kaufmann Publishers Inc., 2012.

\bibitem{d2geo}
B.~Fitz-Gerald, ``D2geo: A framework for computing seismic attributes with
  python,'' 2018, online; available at \url{https://github.com/dfitzgerald3};
  accessed 14-April-2020.

\bibitem{wrona2018seismic}
T.~Wrona, I.~Pan, R.~L. Gawthorpe, and H.~Fossen, ``Seismic facies analysis
  using machine learning,'' \emph{Geophysics}, vol.~83, no.~5, pp. O83--O95,
  2018.

\bibitem{apache-parquet}
D.~Vohra, ``Apache parquet,'' in \emph{Practical Hadoop Ecosystem}.\hskip 1em
  plus 0.5em minus 0.4em\relax Springer, 2016, pp. 325--335.

\bibitem{rapids-cuml}
S.~Raschka, J.~Patterson, and C.~Nolet, ``Machine learning in python: Main
  developments and technology trends in data science, machine learning, and
  artificial intelligence,'' \emph{arXiv preprint arXiv:2002.04803}, 2020.

\bibitem{scikit-learn}
\BIBentryALTinterwordspacing
F.~Pedregosa, G.~Varoquaux, A.~Gramfort, V.~Michel, B.~Thirion, O.~Grisel,
  M.~Blondel, P.~Prettenhofer, R.~Weiss, V.~Dubourg, J.~Vanderplas, A.~Passos,
  D.~Cournapeau, M.~Brucher, M.~Perrot, and {{\'E}}douard Duchesnay,
  ``Scikit-learn: Machine learning in python,'' \emph{Journal of Machine
  Learning Research}, vol.~12, no.~85, pp. 2825--2830, 2011. [Online].
  Available: \url{http://jmlr.org/papers/v12/pedregosa11a.html}
\BIBentrySTDinterwordspacing

\end{thebibliography}

\end{document}